\documentstyle[epsfig]{aipproc}

\newcommand{\etal}{{\it et al.}}
\newcommand{\gray}{$\gamma$-ray}
\newcommand{\grays}{$\gamma$-rays}
\newcommand{\apj}{{\it Ap.J.}}

\begin{document}

\title{Absorption of High Energy Gamma Rays by Interactions\\
with Extragalactic Starlight Photons at High Redshifts}
\author{M.H. Salamon$^*$ and F.W. Stecker$^{\dagger}$}
\address{$^*$Physics Department, University of Utah\\
Salt Lake City, UT 84112\\
$^{\dagger}$Laboratory for High Energy Astrophysics\\
NASA/Goddard Space Flight Center\\
Greenbelt, MD 02271}
\maketitle

\begin{abstract}
We extend earlier calculations of the attenuation suffered by \grays\ during 
their propagation from extragalactic
sources, obtaining new extinction curves for \grays\ down to
10 GeV in energy, from sources up to a redshift of $z=3$.  
\end{abstract}

The recognition that high energy \grays, propagating over cosmological
distances, suffer electron-pair-producing
interactions with photons from the extragalactic background radiation fields dates back
to the 1960s \cite{mhsfws:nis62,mhsfws:gou66,mhsfws:ste69,mhsfws:faz70}.
The reaction $\gamma\gamma\rightarrow
e^{+}e^{-}$ between a \gray\ of energy $E$ and a background
photon of energy $\epsilon$ can occur when 
the center-of-mass square energy $s$ is above threshold,
$s=2E\epsilon(1-\cos\theta)>4m_{e}^{2}c^{4}$, where $\theta$ is the angle between the
two photons' direction vectors.  A \gray\ of energy $E_{\rm TeV}$ TeV therefore
interacts only with
background photons above a threshold 
energy $\epsilon_{\rm thr}\approx 0.3{\rm eV}/E_{\rm TeV}$.
Since the number density of background
photons decreases roughly as a power law in energy, most of the collisions occur near
threshold.  Thus, when estimating the extinction of 1 TeV gamma rays, it is the
density of the infrared background which dominates; at 20 GeV, however, 
only UV photons near the
Lyman limit can act as targets.

This mechanism was recently invoked \cite{mhsfws:ste92}
to explain why many EGRET blazars are not seen
at $\sim$TeV energies by ground-based instruments such as Whipple, in spite of the 
fact that an extrapolation of the EGRET power-law spectra places them above the
sensitivity limit of these ground-based detectors.  
The opacity $\tau$ seen by a \gray\ in its propagation from source to Earth is roughly
$\tau\sim N\sigma_{T} d$, where $N$ is the number of target soft photons above threshold,
$\sigma_{T}$ is the Thompson cross section, and $d$ is the distance to the source.
For sources with redshift $z>0.1$ (corresponding to most of the EGRET blazar sources), based on
estimates of the diffuse IR background \cite{mhsfws:ste97}
the opacity is greater than unity for $\sim$ TeV \grays, making their ground-based detection
unlikely.  

With the advent of a new generation of ground-based instruments with anticipated \gray\ energy
thresholds as low as 20 GeV, and with the likely future launch of GLAST \cite{mhsfws:blo96}
with sensitivity in the range
$\sim$0.01 to 100 GeV, it is important to extend the opacity calculations down to
the lowest relevant \gray\ energies.  Although efforts along these lines have been made
\cite{mhsfws:mad96,mhsfws:ste96a}, 
very recent work on the evolution of star formation rates with
redshift \cite{mhsfws:pei95,mhsfws:fal96} justifies a new and more
detailed calculation of \gray\ opacity.

\begin{figure}
\centerline{\epsfig{file=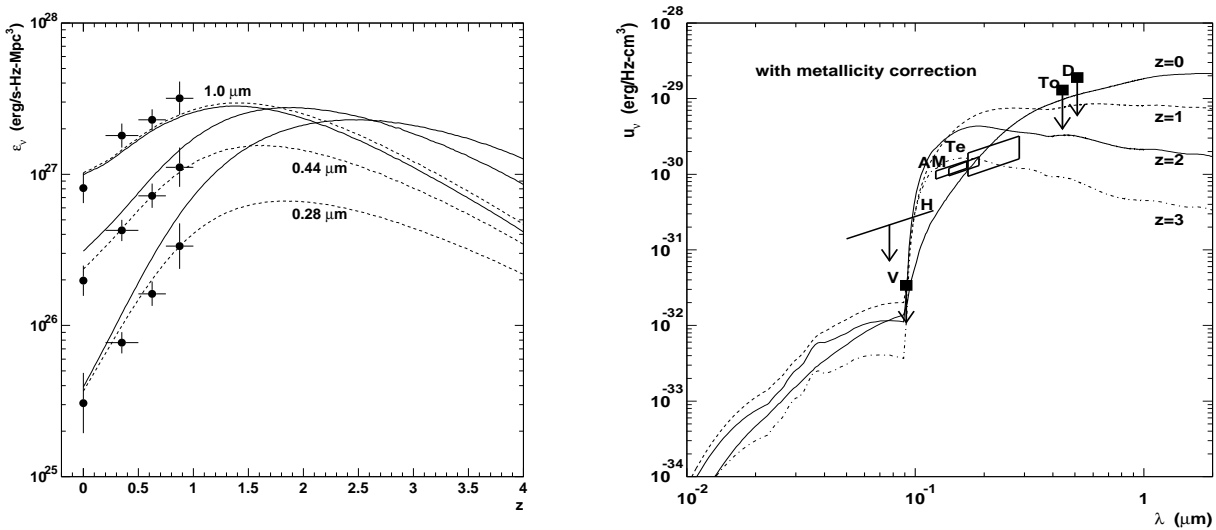,height=3.00in}}
\caption[]{Left: The mean co-moving
emissivity of stellar populations as a function of redshift
for three different wavelengths, with (solid lines) 
and without (dashed lines) our
metallicity correction added.  The dashed lines are essentially a
reproduction of the results of Ref.\cite{mhsfws:fal96}, and 
the observational data
points are from the Canada-French redshift survey 
group \cite{mhsfws:lil96}.  Right: The computed co-moving background 
energy density as a function of wavelength for
several redshifts.  The data points shown are high galactic 
latitude detections or limits at redshift $z=0$: A, 
Ref.\cite{mhsfws:and79}; D, Ref.\cite{mhsfws:dub79}; 
H, Ref.\cite{mhsfws:hol86}; M, Ref.\cite{mhsfws:mar91}; Te, 
Ref.\cite{mhsfws:ten88}; 
To, Ref.\cite{mhsfws:tol83}; V, Ref.\cite{mhsfws:vog95}.
}
\end{figure}

The role played by the extragalactic starlight background
(ESB) in the attenuation of \grays\ from extragalactic sources
is defined in the exact expression for the \gray\ opacity $\tau$,
\begin{equation}
\tau(E_{0},z_{e})=c\int_{0}^{z_{e}}dz\,\frac{dt}{dz}\int_{0}^{2}dx\,\frac{x}{2}
\int_{0}^{\infty}d\nu\,(1+z)^{3}\frac{u_{\nu}(z)}{h\nu}\sigma_{\gamma\gamma}(s),
\label{mhsfws:eq1}
\end{equation}
where $E_{0}$ is the {\it observed} \gray\ energy, $z_{e}$ is the source redshift,
$t(z)$ is the cosmic time,
$x\equiv(1-\cos\theta)$, $\theta$ is the angle between the photons' direction
vectors, $\nu$ is the target photon frequency at redshift $z$, $u_{\nu}(z)$ is
the photon energy density per unit frequency at redshift $z$, and
$\sigma_{\gamma\gamma}$ is the Bethe-Heitler cross section for $\gamma\gamma
\rightarrow e^{+}e^{-}$.

Apart from the uncertainty in cosmological parameters, the only unknown in the above
equation is the ESB energy density $u_{\nu}(z)$.  This can be determined if the
mean emissivity per unit frequency, ${\cal E}_{\nu}(z)$, of starlight from galaxies is known:
\begin{equation}
u_{\nu}(z)=\int_{z}^{z_{\rm max}}dz^{\prime}\,\frac{dt}{dz}
{\cal E}_{\nu^{\prime}}(z^{\prime})e^{-\tau_{\rm cloud}(\nu,z,z^{\prime})}.
\label{mhsfws:eq2}
\end{equation}
Here $z_{\rm max}$ is the redshift for the turn-on of star formation, 
$\nu^{\prime}=\nu(1+z^{\prime})/(1+z)$, and the last factor accounts
for the partial absorption of the starlight
by intervening Ly$\alpha$ clouds during the ESB's
propagation through intergalactic space \cite{mhsfws:shu96}.

The mean emissivity ${\cal E}_{\nu}(z)$ is the total stellar energy output per 
unit volume and frequency, averaged over all galaxies and proto-galaxies, at a given
redshift.  Consider a population of stars all born at the same instant, with an
initial mass function (IMF) $\phi(M)\,dM\propto M^{-\alpha}\,dM$ ($\alpha=2.35$ here).
The total emission $S_{\nu}(T)$ from this population is the integral of 
the spectral energy output of each star (a function of its mass $M$, age
$T$, and to a smaller extent its metallicity 
\cite{mhsfws:wor94,mhsfws:cha96}) weighted 
by the IMF.  As the age $T$ of the population increases,
$S_{\nu}(T)$ becomes redder, due to the shorter lifetimes of the bluer stars.
The mean emissivity ${\cal E}_{\nu}(z)$ is then the
convolution of $S_{\nu}(T)$ with the redshift-dependent stellar formation rate,
$\dot{\rho}_{s}(z)$:
\begin{equation}
{\cal E}_{\nu}(z)={\cal T}_{d,g}(\nu)\int_{z}^{z_{\rm max}}dz^{\prime}\,
\frac{dt}{dz^{\prime}}\dot{\rho}_{s}(z^{\prime})S_{\nu}
\left[T=t(z)-t(z^{\prime})\right]{\cal L}(\nu,z^{\prime}),
\label{mhsfws:eq3}
\end{equation}
where ${\cal T}_{d,g}(\nu)$ is the probability that stellar photons of frequency
$\nu$ will escape absorption by dust and gas in their parent galaxy, and
${\cal L}(\nu,z)$ is a frequency-dependent correction to $S_{\nu}$ which accounts
for the increase in stellar metallicities with decrease in $z$.

Figure 1 shows our results for ${\cal E}_{\nu}(z)$
(Eq.3) and $u_{\nu}(z)$ (Eq.2).  For $S_{\nu}(T)$
we have used the population synthesis models of 
Refs.\cite{mhsfws:bru93,mhsfws:cha91,mhsfws:cha96}; 
for ${\cal L}(\nu,z)$ we have constructed an empirical correction
function based on the work of Ref.\cite{mhsfws:wor94}; the star formation rate
$\dot{\rho}_{s}(z)$ comes from the beautiful analysis of 
Refs.\cite{mhsfws:pei95,mhsfws:fal93}.  
(See Ref.\cite{mhsfws:sal97} for more details.)

With Eq.1, the ESB opacity to \grays\ is calculated, and shown
in Fig.2, both with and without the metallicity correction
function ${\cal L}$ included.  Given the uncertainties associated with 
${\cal L}$ \cite{mhsfws:sal97}, the true opacities likely lie somewhere between
the two sets of curves.

\begin{figure}
\centerline{\epsfig{file=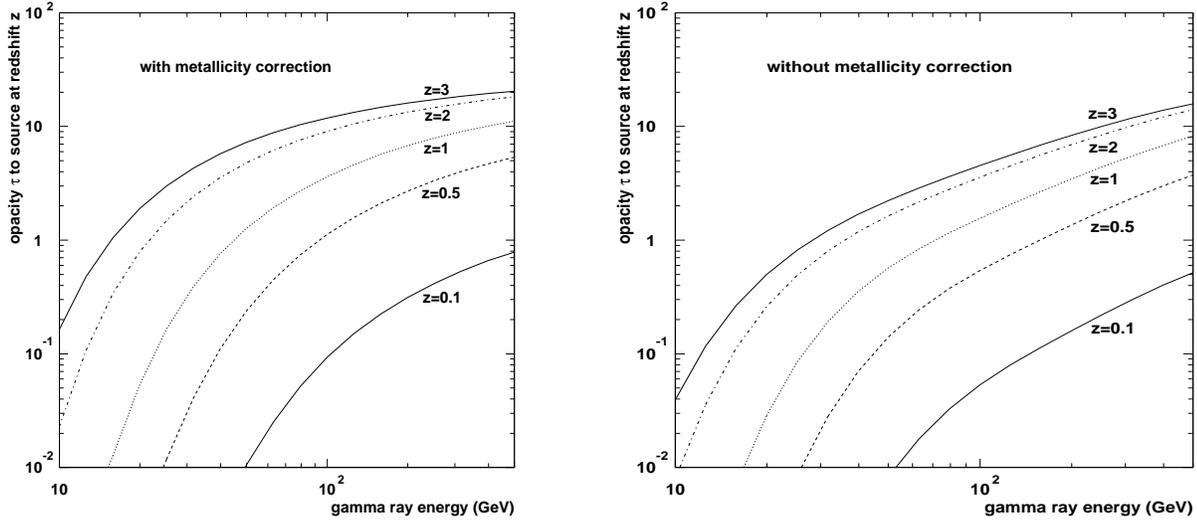,height=3.0in}}
\caption[]{The opacity $\tau$ of the ESB to 
\grays\ as a function of \gray\ energy and source redshift $z$. Left:
Curves calculated with the metallicity correction included.  Right: Curves
calculated without metallicity correction.
The truth likely lies between these two sets
of curves.  We note that the opacities obtained here are independent of the value
assumed for Hubble's constant (see Ref.\cite{mhsfws:sal97} for details).}
\end{figure}

Figure 3 shows the effect of the ESB on \gray\ propagation
from several blazars.  Note that the spectral cutoffs occur at lower energies
for blazars at higher redshifts, a distinctive signature which can discriminate
this cutoff mechanism from intrinsic (intrasource) cutoff mechanisms.  Also note
that there is essentially no attenuation below 10 GeV, due to the sharp break in
the energy density above the Lyman limit (Fig.1).
Figure 3 also shows the beginning of the extinction of that component
of the extragalactic \gray\ background above 20 GeV that is due to unresolved
blazars \cite{mhsfws:sal96}; this is compared with recent EGRET measurements
of the extragalactic \gray\ background \cite{mhsfws:sre97}.

\begin{figure}
\centerline{\epsfig{file=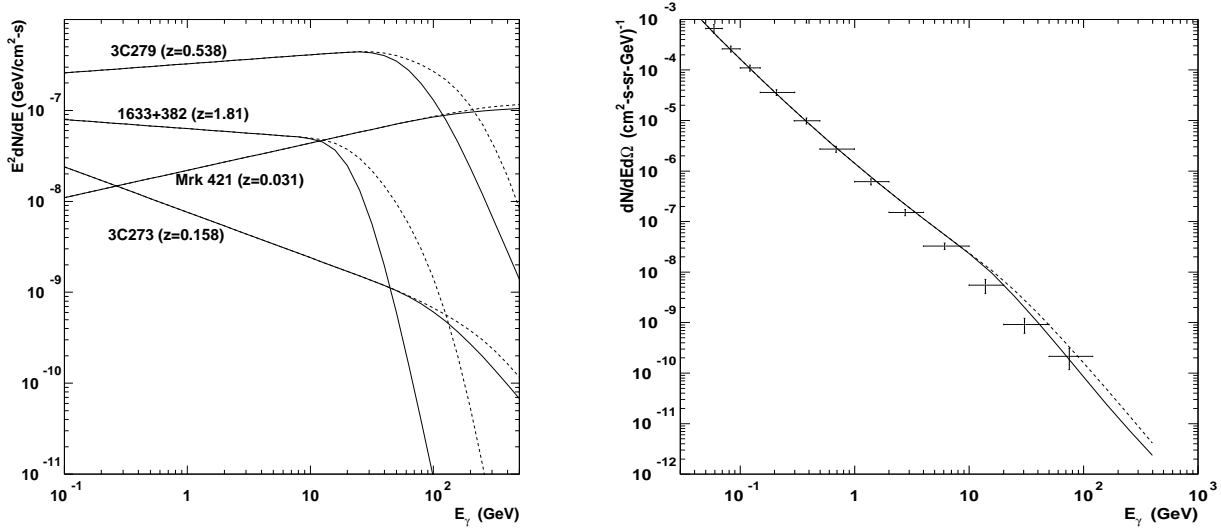,height=3.0in}}
\caption[]{Left: The attenuated power-law spectra of four prominent blazars.  The
solid (dashed) curves are calculated with (without) the metallicity correction
factor.  Right: Extragalactic \gray\ background spectrum from unresolved blazars,
calculated for the EGRET point source sensitivity of $10^{-7}$ cm$^{-2}$s$^{-1}$;
solid (dashed) line includes (does not include) metallicity correction, and data
points are from EGRET \cite{mhsfws:sre97}.
}
\end{figure}

\noindent
{\bf Acknowledgements:} We thank M. Fall, M. Malkan,
Y.C. Pei, P. Sreekumar, G. Worthey, and N. Wright helpful conversations
and advice.

\end{document}